\documentclass[conference]{IEEEtran}
\IEEEoverridecommandlockouts
\usepackage{graphicx}
\graphicspath{{figures/}}
\usepackage[caption=false,font=footnotesize]{subfig}
\usepackage{float}

\usepackage{cite}

\usepackage{amsmath,amssymb,amsfonts}
\usepackage{textcomp}
\usepackage{xcolor}
\usepackage{romannum}
\usepackage[linesnumbered,ruled,vlined]{algorithm2e}

\makeatletter

\def\ps@IEEEtitlepagestyle{%
    \def\@oddfoot{\mycopyrightnotice}%
    \def\@evenfoot{}%
}
\def\mycopyrightnotice{%
    {\footnotesize  978-1-7281-7366-5/20/\$31.00 \textcopyright2020 IEEE\hfill}
    \gdef\mycopyrightnotice{}
}

\makeatletter
\newcommand{\removelatexerror}{\let\@latex@error\@gobble}
\makeatother
\usepackage{algpseudocode}

\SetKwInput{KwInput}{START}                
\SetKwInput{KwOutput}{Output}              

\def\BibTeX{{\rm B\kern-.05em{\sc i\kern-.025em b}\kern-.08em
    T\kern-.1667em\lower.7ex\hbox{E}\kern-.125emX}}

\author{\IEEEauthorblockN{Sharmin Akter$^{1}$, Mohammad Shahriar Rahman$^{2}$ and Nafees Mansoor$^{3}$}
\IEEEauthorblockA{Department of Computer Science and Engineering, University of Liberal Arts Bangladesh}
Email:{sharmin.akter2.cse@ulab.edu.bd$^{1}$, shahriar.rahman@ulab.edu.bd$^{2}$, nafees@ieee.org$^{3}$}
}



\makeatletter
\newcommand*\titleheader[1]{\gdef\@titleheader{#1}}
\AtBeginDocument{%
  \let\st@red@title\@title%
  \def\@title{%
    \bgroup\normalfont\large\raggedright\@titleheader\par\egroup
    \vskip1.5em\st@red@title}
}
\makeatother

\title{An Efficient Routing Protocol for Secured Communication in Cognitive Radio Sensor Networks}

\titleheader{2020 IEEE Region 10 Symposium (TENSYMP), 5-7 June 2020, Dhaka, Bangladesh
}

\begin{document}
\maketitle

\begin{abstract}
This paper introduces an efficient reactive routing protocol considering the mobility and the reliability of a node in Cognitive Radio Sensor Networks (CRSNs). The proposed protocol accommodates the dynamic behavior of the spectrum availability and selects a stable transmission path from a source node to the destination. Outlined as a weighted graph problem, the proposed protocol measures the weight for an edge the measuring the mobility patterns of the nodes and channel availability. Furthermore, the mobility pattern of a node is defined in the proposed routing protocol from the viewpoint of distance, speed, direction, and node's reliability. Besides, the spectrum awareness in the proposed protocol is measured over the number of shared common channels and the channel quality. It is anticipated that the proposed protocol shows efficient routing performance by selecting stable and secured paths from source to destination. Simulation is carried out to assess the performance of the protocol where it is witnessed that the proposed routing protocol outperforms existing ones.

\end{abstract}


\section{Introduction}

With latest advancement of the wireless communication, Wireless Sensor Networks (WSN) are the devices consisting of large number of low powered battery sensors which has the ability to sense, transmit data and relay to other sink nodes. It has earned profound popularity in wide range of sensor based application in different wireless communication areas. It is designed for fixed and licensee free Spectrum. Hence, there results in excessive interference occur during sharing same channel in data dissemination. As a result, packet loss or network failure is endangered due to this interference. So efficient utilization of spectrum has resulted in a challenge. To overcome this challenge, CR technology has emerged as a prominent technology which utilize the radio spectrum efficiently in a opportunistic manner \cite{sharma} and thus improves spectrum efficiency. By providing CRN technology to WSN which prompts to Cognitive Radio Sensor Network (CRSN), WSN nodes capable to get the facilities of the CRNs and will be able to exploit the utilized spectrum intelligently.

In this paper a new trust based secure routing mechanism is introduced for route discovery and selection process considering spectrum scarcity and dynamic mobility pattern of the spectrum band. Here, mobility, distance, link stability are considered both in spectrum sensing and data transmission \cite{akter}. A clustering mechanism is also presented in this paper for discovering stable route for faster data delivery.

Many protocols of CRSNs are proposed in the previous research work. Based on the limited energy of the routing protocol has been proposed in \cite{bukhari} but it does not consider spectrum mobility and lacks secure communication. For assuring secure communication detection of malicious node algorithm is proposed in \cite{dmn} which improves network performance considering distrust value. A cluster based scheduling algorithm has been proposed in \cite{idoudi} which considers sensor energy and the collision of the CR nodes in the clusters. Based on the energy consumption in CR nodes, a cluster based routing is demonstrated in \cite{ren} which reduces energy consumption and also improves free channel utilization. By implementing dijkstra algorithm an energy efficiency routing protocol in \cite{tony} which is optimized to avoid interference of PU user in channel switching. Considering mobility, distance, velocity of CR nodes routing protocol has been proposed in \cite{mpbrp}. Then, based on velocity and location of the nodes the presented routing protocol \cite{pgrp} does not consider spectrum availability, link quality and reliability. Based on the location and density, a software based routing protocol is presented in \cite{sdgr} which also does not utilize the spectrum bands and does not discuss about the link quality, collision of CR nodes and reliability.

The rest of the article is organized as follows. Section II depicts a network model of the proposed routing protocol. Section III discusses the proposed protocol and the routing mechanism. Then simulation results and performance evaluation are discussed in section IV. Finally, the paper concludes with the conclusion and future work in section V.

\section{Network Model}
In this proposed routing protocol, each CR in the network has the ability to sense the free spectrum bands associated with two transceivers equipment for control and data transmission. It is assumed that the transmission range is equivalent to all nodes and each CR node is capable to compute the NHDF (Next-hop Determination Factor) \cite{akter} and IF (Intrusion Detection Factor). To distinguish the licensed spectrum of PU's, SU observes its nearby radio transmission. The proposed spectrum aware clustering mechanism separates the network transmission system into legitimate groups. By propagating the neighbour nodes' list after the neighbor discovery is executed, ACL exchange is accomplished among 1 -hop neighbors. To diminish the number of channels and furthermore development of clusters with maximum common channels, a cluster formation mechanism, earlier published by the author, named RARE is utilized in the proposed routing protocol\cite{rare}.

 Here, both CR nodes anticipate link quality for estimating the stability of the transmission  between  two  nodes  in  the  network.  The  source node chooses the following neighboring node with a minimum delay  which  has  better  link  quality.  Here,  each  node  knows about traffic density as having a similar channel by the node causes  congestion. For increasing the security a concept of blockchain consensus mechanism flowchart is proposed in Fig. 1 for detecting any abnormal behavior of vehicles or any  internal or  external  attack for  example, DoS,  Black-hole,  sticking,  and  so  on which  keep track of the trust level and forwards safe messages for transmission though discarding the malicious node to achieve stable and reliable  communication  in  the  same  network  transmission.

\section{Proposed Routing Protocol}

In this routing protocol, a routing metric namely $NHDF$ is presented \cite{akter}. Here, total cumulative $N$ is determined with respect to routing path which is estimated as follows,

\begin{equation}
     N _{i,j}= \frac{(\frac{\xi_T}{\delta_{i,j}^E})^{C_n}}{IF}
\end{equation}

Here, $\xi_T$ is transmitting weight value  dependent on transmission range ($\Phi_t$), displacement $\tau_v$, link delay $\delta^L_p$ and speed (s).

\begin{equation}
    \xi_T = \frac{\Phi_t}{\tau_v * \delta^L_P * s}
\end{equation}

At the time $T2$ the displacement ($\tau_v$) between two neighbor vehicular nodes is calculated below,

\begin{equation}
    \tau_v = d\theta
\end{equation}

At first during broadcasting RREQ distance ($d$) is estimated measuring the RSSI value which is demonstrated as follows,

\begin{equation}
    d = 10^{(\frac{\kappa - 20log(\frac{4\pi l_0}{\upsilon})}{10\omega})}\ l_0
\end{equation}

Here,  where k is considered as path loss,  $\omega$  is the signal loss exponent,  $\upsilon$ is the wavelength of the received signal and $l_0$ is reference distance.

During this data dissemination the sensor node's speed (s) can be expressed as follows,

\begin{equation}
s = \frac{\sqrt{(\alpha_r - \gamma_r)^2 + (\alpha_t - \gamma_t)^2 }}{(T_2 + \Delta ) - T_1}
\end{equation}

where, $T_1$ and $T_2$ is the received and sending timestamp respectively alongside present co-ordinates by waiting for a random back-off time. Here, $\psi_r(\alpha_r, \gamma_r)$ is denoted as the co-ordinate of sensor nodes during receiving and $\psi_s(\alpha_t, \gamma_t)$ is characterized as the co-ordinate of the vehicular node during sending. $\Delta$ is denoted as transmission time. 
Then, direction between neighbor node and the destination node is estimated as follows,

\begin{equation}
    \theta = \cos^{-1}(\frac{\overrightarrow{\psi_r  \psi_s}\overrightarrow{\varphi_r \varphi_s}}{||\overrightarrow{\psi_r \psi_s}||\times ||\overrightarrow{\varphi_r \varphi_s}||})
\end{equation}

where, $\varphi_r(\alpha_s, \gamma_s)$ is the destination position during receiving data packets and $\varphi_s(\alpha_q, \gamma_q)$ is  while sending data packets.

Three kinds of delays such as back-off delay, queuing delay and switching delay are considered in this routing protocol between the the intermediate nodes. If the quantity of neighboring sensor node is resolved as $V_i$, $RT_i$ is indicated as data rate of $N_i$ and \textit{S} is the size of the data, the queuing delay $(\delta_L^N)$ of $N_i$ can be evaluated as follows, 

\begin{equation}
    \delta_i^N=\frac{SV_i}{RT_i}
\end{equation}

Sensor nodes utilize random back-off time as various sensor nodes may utilize a similar channel to avoid collision during same channel sharing. By the below equation the back-off delay can be estimated,

\begin{equation}
    \delta_i^K= \frac{1}{(1 - b_c)(1 - (1-b_c)^{V_i-1})}\ z 
\end{equation}

here, $b_c$ is recognized as the probability of collision and $V_i$ is the neighboring sensor node on a channel $CH_i$ and $z$ is the window size.

Hence, during forwarding the data packets if a node $N_i$ is required to switch from channel p to divert q in its channel-group during sending the message to next-hop $N_j$, the calculated channel switching delay is characterized as follows, 

\begin{equation}
    \delta_{i,j}^M = a* |p - q |
\end{equation}

where $a$ is positive real number and for specific step size $a$ is considered as the tuning delay of two neighboring channels. For step size 10MHz, $a$ is denoted as 10ms [14]. 

Thus, the delay of the link $(\delta_{i,j}^E)$ that connects $N_i$ and $N_j$ is calculated using Equation 7, 8 and 9 as below,

\begin{equation}
    \delta_{i,j}^E= \delta_{i,j}^M+ \delta_i^N+ \delta_i^K
\end{equation}

The Intruder Determination Factor ($IF_{CV_i}^{CN_j}$) of the neighboring node ($CN_j$) for any sensor node ($CV_i$) can be evaluated as follows, 

\begin{equation}
    IF_{CV_i}^{CN_j} = e^{RN}
\end{equation}

here, $RN$ indicates the report number during each question or suspect by vehicular nodes during transmission. 

\begin{table}[ht]
\caption{Symbols used in the proposed routing protocol}
\label{tab:title}
\begin{tabular}{ll}
    \hline
      \textbf{Symbols} & \textbf{Description} \\
      \hline
      $CH$ & Cluster head \\
      $CS_i$ & Neighbors of channel set \\
      $V_j$ & Joining node \\
    $RREQ$ & Route Request during data transmission \\
     $CN_j$ & Neighbors of cluster set \\
    $CV_i$ & Any sensor node in cluster \\
     $MCV_i$ & Malicious behavior of any vehicle $CV_i$ \\
     $IF$ & Intrusion detection factor \\
    $QM_s$ & 50\% malicious suspect or query by any $CV_i$  \\
     $RN$ & Report number \\
     $BV_i$ & Behaviour of $CV_i$ \\
      \hline
    \end{tabular}
\end{table}

\begin{figure}[htp]
    \centering
    \includegraphics[]{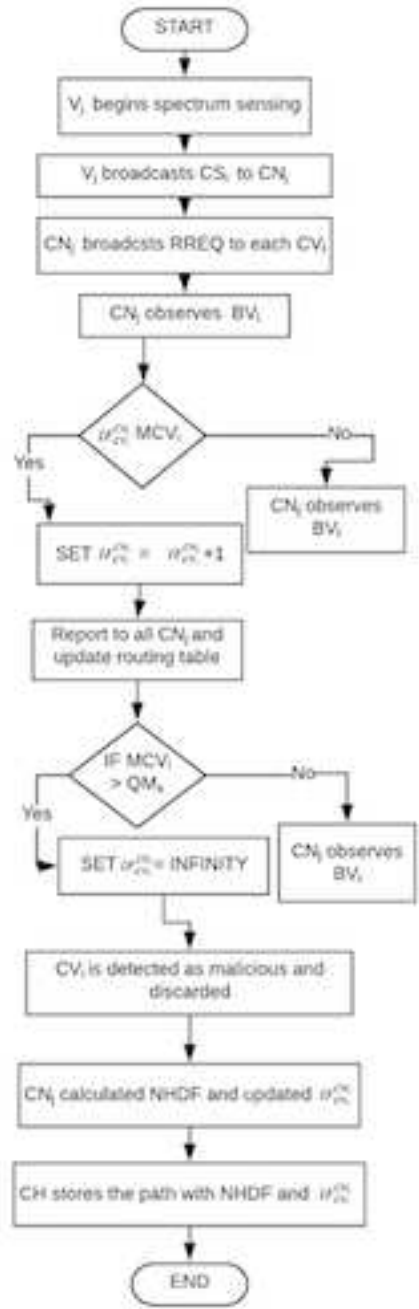}
    \caption{Flowchart of the Proposed Routing Protocol}
    \label{fig:vanet}
\end{figure}

\subsection{Route Discovery and Selection Process}

In the proposed routing protocol, at first the joining node of a cluster $V_j$ starts spectrum sensing. Then the $V_j$ broadcasts accessible channel set $CS_i$ to every neighbor of cluster set $CN_j$. Then RREQ is broadcast by $CN_j$ While broadcasting RREQ the neighboring or relay nodes observe the behavior of $CV_i$ and then estimate $IF$ value by implementing intruder detection mechanism demonstrated in Fig. 1. Then $CN_j$ updates the routing table with $IF$ value. When the destination node exists in the neighboring list, the neighboring node estimates the $NDHF$ value by equation 1. After that cluster head $CH$ initiates a RREP with $NHDF$ and $IF$ with the routing path to the previous generated RREQ message. This is a continuous process until the neighbor node receives
the RREP message and all possible routing paths are discovered. Lastly, the $NHDF$ and $IF$ values are stored in in routing path array calculated by the source node.

\subsection{Reliable Data Dissemination Mechanism}

Here, in the intruder detection mechanism every node joins in the blockchain network and gathers information from the neighboring sensor nodes. During sending RREQ every neighboring vehicle $CN_j$ observes the behavior of any sensor node $CV_i$ in the blockchain network. When the neighboring nodes find any report or query or  suspect of showing abnormal behavior of $V_i$ , $IF_{CV_i}^{CN_j}$ increments where the value of $IF_{CV_i}^{CN_j}$ is initially is set to 1 and report to the all neighboring nodes in the blockchain network. Then the neighboring node updates the routing table with new $IF_{CV_i}^{CN_j}$ value. $CN_j$ checks if more than $50\%$ query or suspect of malicious behavior exist, the $IF_{CV_i}^{CN_j}$ is set as INFINITY which detects $CV_i$ as malicious node  and sends warning message to the all the neighboring nodes in the network. Finally, the entry of node $CV_i$ is updated in blacklist and the malicious node $CV_i$ is discarded from the network. Thus consequently we can monitor all the vehicles' present trustworthiness.

\begin{figure*}[t]
\centering
\mbox{\subfloat[Comparison in terms of End-to-End Delay]{\label{delay}\includegraphics[width=0.33\linewidth]{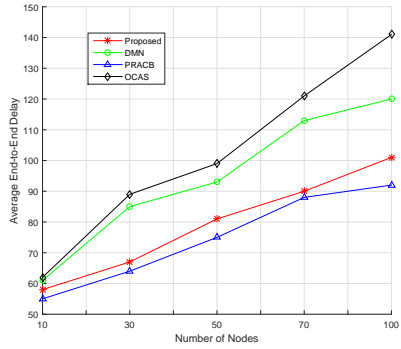}}}
\mbox{\subfloat[Comparison in terms of Throughput]{\label{throughput}\includegraphics[width=.33\linewidth]{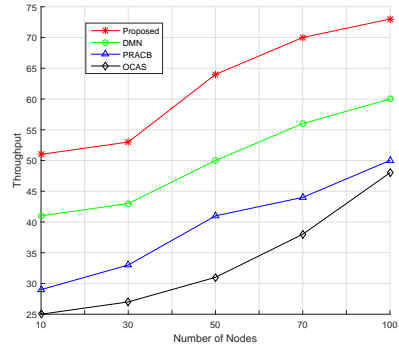}}}
\caption{Performance comparisons of the proposed routing protocol with other protocols.}
\label{simu}
\end{figure*}

\section{Simulation Results}

The performance of the proposed routing protocol is evaluated by conducting simulation on NS2 which is a discrete event simulator. The network setup is associated with 10 clusters consisting of 400 sensor nodes. PUs and SUs are randomized position in this network where other configurations are listed in the Table II. To analyze the performance a comparison study has been conducted with the existing routing protocols namely OCAS \cite{sharma}, PRACB \cite{bukhari} and DMN \cite{dmn} considering throughput and end-to-end delay.

\begin{table}[ht]
\caption{Simulation Environments}
\label{tab:simu}
\begin{center}

\begin{tabular}{ll}
    \hline
    \textbf{Parameters} & \textbf{Value} \\
      
     \hline
      Number of vehicular nodes & 10,30,50,70,100\\
      Simulation area & 4000$m^2$ \\
      Run time & 120 s \\
      Packet size & 256 bytes \\
      Traffic type & CBR \\
      CBR & 5 packets/s \\
      Queue type & Drop-tail \\
      Transmission range & 500 m \\
      Propagation model & Two-Ray Ground \\
      Mac Layer Protocol & 802.15.6p \\
      Number of malicious nodes & 5  \\
      speed of vehicles & 5 m/s \\
      \hline
    \end{tabular}
  \end{center}
\end{table}

Fig.  1(a)  shows the comparing results of the proposed routing protocol with other existing routing protocols OCAS, PRACB and DMN in terms of End-to-End delay. Here, the delay increases as the density of the network increases with the number of nodes and this requires more time which results in more delay. Here, PRACB performs better as it lacks of considering malicious or any abnormal behavior of node during data transmission. On the other hand, the proposed routing protocol which shows little bit higher delay rather than PRACB, considers reliability and observes malicious behavior for intruder detection. That is why the trusted and reliable path selected by proposed routing protocol is longer than the shortest one during data transmission.

Fig.  1(b)  depicts  that  our proposed routing protocol outperforms than other existing routing protocol OCAS, PRACB and DMN. Here, throughput increases with the number of nodes as more sensor nodes are connected in the network for packet transmission which causes less packet drop or network failure. The proposed routing protocol performs better as it considers mobility as routing metric and also estimates stable route for both in spectrum sensing and data transmission.

\section{Conclusion and Future Works}

A secured reactive routing protocol for CRSN is presented in this paper. The double-folded dynamic behavior of the network is  considered in the proposed protocol where autonomous movements of the sensors and changing spectrum availability are measured. From the simulation results, it is observed that the proposed routing protocol performs better than other recently introduced protocols. This study will lead to the further development of the routing protocol where the energy efficiency of the network will be addressed.


\begin{thebibliography}{}


\bibitem{sharma}{
Sharma, K.K. and Trivedi, A., 2015, April. An Opportunistic channel access scheme with channel ordering for cognitive radio network. In 2015 Fifth International Conference on Communication Systems and Network Technologies (pp. 354-358). IEEE.
}

\bibitem{Mitola}{Mitola, J. and Maguire, G.Q., 1999. Cognitive radio: making software radios more personal. IEEE personal communications, 6(4), pp.13-18.}

\bibitem{Nafees_Survey}{Mansoor, N., Islam, A.M., Zareei, M., Baharun, S., Wakabayashi, T. and Komaki, S., 2015. Cognitive radio ad-hoc network architectures: a survey. Wireless Personal Communications, 81(3), pp.1117-1142.}

\bibitem{idoudi}{Idoudi, H., Mabrouk, O., Minet, P. and Saidane, L.A., 2019. Cluster-based scheduling for cognitive radio sensor networks. Journal of Ambient Intelligence and Humanized Computing, 10(2), pp.477-489.
}

\bibitem{ren}{Ren, J., Zhang, Y., Zhang, N., Zhang, D. and Shen, X., 2016. Dynamic channel access to improve energy efficiency in cognitive radio sensor networks. IEEE Transactions on Wireless Communications, 15(5), pp.3143-3156.
}

\bibitem{bukhari}{Bukhari, S.H.R., Siraj, S. and Rehmani, M.H., 2016. PRACB: A novel channel bonding algorithm for cognitive radio sensor networks. IEEE Access, 4, pp.6950-6963.}

\bibitem{pgrp}{Karimi, R. and Shokrollahi, S., 2018. PGRP: Predictive geographic routing protocol for VANETs. Computer Networks, 141, pp.67-81.}

\bibitem{tony}{Cladia, A.T. and Rajavel, S.E., 2018, December. Optimizing Spectrum Sensing For Energy Efficient Cognitive Radio Sensor Networks. In 2018 International Conference on Smart Systems and Inventive Technology (ICSSIT) (pp. 333-338). IEEE. }

\bibitem{sdgr}{Ji, X., Yu, H., Fan, G. and Fu, W., 2016, December. SDGR: An SDN-based geographic routing protocol for VANET. In 2016 IEEE International Conference on Internet of Things (iThings) and IEEE Green Computing and Communications (GreenCom) and IEEE Cyber, Physical and Social Computing (CPSCom) and IEEE Smart Data (SmartData) (pp. 276-281). IEEE.}

\bibitem{mpbrp}{
Ye, M., Guan, L. and Quddus, M., 2019, April. Mpbrp-mobility prediction based routing protocol in vanets. In 2019 International Conference on Advanced Communication Technologies and Networking (CommNet) (pp. 1-7). IEEE.}

\bibitem{dmn}{Khan, U., Agrawal, S. and Silakari, S., 2015. Detection of malicious nodes (dmn) in vehicular ad-hoc networks. Procedia computer science, 46(9), pp.965-972.}

\bibitem{dmv}{Daeinabi, A. and Rahbar, A.G., 2013. Detection of malicious vehicles (DMV) through monitoring in Vehicular Ad-Hoc Networks. Multimedia tools and applications, 66(2), pp.325-338.}

\bibitem{rare}{Mansoor, N., Islam, A.M., Zareei, M. and Vargas-Rosales, C., 2018. RARE: A spectrum aware cross-layer MAC protocol for cognitive radio ad-hoc networks. IEEE Access, 6, pp.22210-22227.}

\bibitem{akter}{Akter, S. and Mansoor, N., 2020, May. A Spectrum Aware Mobility Pattern Based Routing Protocol for CR-VANETs. In 2020 IEEE Wireless Communications and Networking Conference (WCNC2020). IEEE. }

\end{thebibliography}
\end{document}